\documentclass[preprint]{aastex}

\begin{document}
\newcommand{\srca}{GX~17$+$2}
\newcommand{\srcb}{4U~1705$-$44}
\newcommand{\srcc}{4U~1728$-$34}

\title{Column Densities Towards Three Bursting Low-Mass X-ray Binaries from 
High Resolution X-ray Spectroscopy}

\author{Patricia Wroblewski, Tolga G\"uver, and Feryal \"Ozel}

\affil{University of Arizona, Departments of Astronomy and Physics, 
933 N. Cherry Ave., Tucson, AZ 85721}

\begin{abstract}

We measured the galactic hydrogen column densities to the neutron-star
binaries GX~17$+$2, 4U~1705$-$44, and 4U~1728$-$34 by modeling the Mg
and Si absorption edges found in high-resolution X-ray spectra
obtained by the Chandra X-ray Observatory. We found for GX~17$+$2,
$N_{\rm H} = (2.38 \pm 0.12) \times 10^{22}$~cm$^{-2}$, for
4U~1705$-$44, $N_{\rm H} = (2.44 \pm 0.09) \times 10^{22}$~cm$^{-2}$,
and for 4U~1728$-$34, $N_{\rm H} = (2.49 \pm 0.14) \times
10^{22}$~cm$^{-2}$. These values are in reasonable agreement with the
hydrogen column densities inferred earlier from modeling of the
continuum spectra of the sources. Our results can be used to constrain
the uncertainties of model parameters of the X-ray spectra of these
sources that are correlated to the uncertainties of the hydrogen
column density. In the case of continuum spectra obtained during
thermonuclear X-ray bursts, they will significantly reduce the
uncertainties in the spectroscopically measured masses and radii of
the neutron stars.

\end{abstract}


\section{Introduction}

Low-mass X-ray binaries (LMXBs) that show thermonuclear
bursts are ideal targets for measuring the masses and radii of neutron
stars. During these short-lived flashes, the flux from the stellar
surface dominates the persistent luminosity by up to two orders
of magnitude and can often reach the local Eddington limit. Owing to
their low magnetic fields, inhomogeneities in the surface emission are
neither observed (e.g., Galloway et al.\ 2003), nor theoretically
expected, during the peak of the bursts.  Modeling time resolved X-ray
spectra during these bursts can, thus, lead to an accurate measurement
of either the gravitational redshift, or, if the distance to the
source is also known, to a measurement of the mass and the radius of
neutron star (van Paradijs 1978; Damen et al.\ 1990; Lewin, van
Paradijs, \& Taam 1993; \"Ozel 2006; \"Ozel, G\"uver, \& Psaltis
2008).

Accurate measurements of column densities towards the bursting LMXBs
are important for two reasons. First, X-ray spectra are attenuated by
the interstellar medium (ISM) due to photoelectric absorption and
scattering by dust grains. When X-ray spectra, particularly in the
soft X-ray band (0.1-10~keV), are used to measure the effective
temperature of the neutron star, $T_{\rm eff}$, and its radius R, this
extinction affects the modeling of continuum spectra, the inferred
temperature, and, thus, the measured stellar radius. Because of the $R
\sim T_{\rm eff}^{-2}$ scaling of the measured radius, uncertainties in the
column density can give rise to large errors in radius. The spectral
parameters are indeed very sensitive to the assumed or allowed values
of the hydrogen column density. We show in Figure~1 an example of the
correlated uncertainties between the temperature and the column
density for the X-ray spectrum of 4U~1728$-$34 taken with the Rossi
X-ray Timing Explorer Proportional Counter Array during a
thermonuclear burst. Such correlations alone can lead to $\gtrsim
15\%$ uncertainties in stellar radii.

Second, measuring the column of interstellar extinction towards an
X-ray source facilitates the use of a novel distance determination
technique based on red clump stars along the line of sight to the
source. The red clump stars are excellent infrared standard candles
(Paczynski \& Stanek 1998; Lopez-Corredoira et al.\ 2002) that can be
used to measure the run of reddening with distance. This method
establishes a distance ladder in the galaxy, which can then be
compared to the extinction of X-ray sources that is determined by
their high resolution spectra to determine their distances.

An independent measurement of the X-ray column density through a
grating observation, where the photoelectric absorption edges of each
element within the observational window can be determined, is valuable
for correctly computing the spectral parameters of these binary
systems. This is a method that has been previously applied to LMXBs by
Juett, Schulz, \& Chakrabarty (2004) and Juett et al. (2006), who
measured the O, Ne, and Fe edges in their spectra obtained with the
Chandra High Energy Transmission Grating (HETG) and XMM-Newton. It has
also been employed in the case of Anomalous X-ray pulsars by Durant \&
van Kerkwijk (2006).

In this paper, we analyze Chandra HETG spectra taken during numerous
observations of GX~17$+$2, 4U~1705$-$44, and 4U~1728$-$34.  These
binaries do not have previously measured hydrogen column densities
from absorption edges in high-resolution X-ray grating spectra. We
report the equivalent column densities for each of the measurable
edges, investigate any possible systematic uncertainties, and compare
our findings with column densities obtained previously from continuum
spectra for these sources.

All three sources have shown frequent thermonuclear (Type-I) X-ray
bursts, with a total number as high as 106 in the case of 4U~1728$-$34
(Galloway et al.\ 2008). A large number of these bursts (e.g., 69 for
the last source) also showed characteristic photospheric radius
expansion properties, indicating that the local Eddington limit has
been reached. Thus, they are ultimately good candidates for the
measurement of the neutron star mass and radius. Given their locations
in the galactic plane, extinction towards these sources can also be
used to study the properties of the ISM in the Galaxy.

\section{Observations and Data Reduction}

The low-mass X-ray binaries GX~17$+$2, 4U~1705$-$44, and 4U~1728$-$34
were observed multiple times with the High Energy Transmission Grating
Spectrometer (HETG) onboard the Chandra X-ray Observatory. Our study
includes a total of nine archival observations of these three sources.
The HETG consists of the Medium Energy Gratings (MEG), with a
$2.5-31~{\rm \AA}$ wavelength range and a resolution of $\delta
\lambda = 0.023~{\rm \AA}$ (for the first-order spectra), and the High
Energy Gratings (HEG),with a $1.2-15~{\rm \AA}$ range and a resolution
of $\delta \lambda = 0.012~{\rm \AA}$.

The three observations of GX~17$+$2, with 30.18~ks, 23.68~ks, and
24.08~ks durations, were all taken in the continuous clocking (CC)
mode between 2004 and 2006. Both observations of 4U~1705$-$44 were
taken in the timed exposure (TE) mode, for a total exposure of
25.13~ks in 2001 and 27.25~ks in 2005. 4U~1728$-$34 has been observed
four times, with exposure times of 10~ks (in 2002), 151.84~ks (in
2006), 49.49~ks (in 2006), and 39.71 (in 2006). The first observation
in 2002 was in TE mode, while all the subsequent observations have
been performed in the CC mode. We list in Table~1 the observations
included in our study, together with the average count rate in each
observation.

The spectral data were reduced following standard CIAO
threads\footnotemark[1] using the CIAO version 4.0.1 and CALDB
3.4.3. Different threads were used for continuous clock and timed
event modes due to the variations in the calibration. For observations
performed in a timed exposure mode, the high fluxes of the sources can
cause a pile-up in the zeroth order (undispersed) images. This can
give rise to problems in determining the exact zero-order position of
the source, which can ultimately shift the wavelength scale of the
grating spectrum. For this reason, we have made use of the
findzo.sl\footnotemark[2] files, which use the MEG arms and the
frame-shift streak to find the best fit zero-order
position. Additionally, for both types of observations, we excluded
any observed Type-I bursts, which have different continuum spectra and
may also have local heavy-element features.

Detector response files (ARFs) were created for the $+1$ and $-1$ MEG
and HEG spectra for both the TE mode and the CC mode observations.
Note that the ARFs include the correction for any time-dependent
change due to a contaminant on ACIS.  For the TE mode observations,
the $+1$ and $-1$ orders were combined together for both the MEG and
the HEG spectra. However, two concerns rendered the MEG $+1$ order of
the CC mode observations unusable. First, there is an instrumental
feature in the O edge on the MEG $+1$ side (Juett, A., Private
Communication). Second, all the observations we studied were carried
out with either a $0.33^\prime$ or $0.1667^\prime$ offset, which
caused the wavelength of the Mg edge to fall in or near chip gaps. In
the CC mode, this resulted in a complete reduction of the effective
area at that wavelength, while in the TE mode, there were sufficient
counts that were included in the analysis. Figure~\ref{effarea} shows
a sample observation from 4U~1728$-$34, where the effective areas of
each grating and order can be seen. Finally, all MEG data were binned
by a factor of 4 and the HEG data were binned by a factor of 8 to
increase the statistics while achieving the same energy resolution on
both gratings.

\section{Spectral Analysis}

We originally sought to measure the edges of neutral Mg, Si, Ne, O,
and Fe in the HETG spectra of GX~17$+$2, 4U~1705$-$44, and
4U~1728$-$34. However, the source count rates were typically very low
at long wavelengths, leading to negligible flux around the O and Fe
edges for all the sources. Near the Ne edge, at a wavelength of
$14.31~{\rm \AA}$ (Juett et al.\ 2006), the continuum was detectable
but there were not enough counts for any of the sources to yield
statistically significant results. We thus considered only the Mg and
Si edges in our analyses.

We focused on small wavelength regions around the Mg and Si edges and
assumed that the continuum spectrum of each source could be fit with a
power-law in each of these regions, $F^{\rm c}_\lambda \propto
(\lambda/\lambda_{\rm edge})^\alpha$, with a break at the absorption
edge of interest. We used fixed edge wavelengths: $\lambda_{\rm Mg}=
9.5~{\rm \AA}$ for the location of the Mg edge and $\lambda_{\rm
Si}=6.72~{\rm \AA}$ (Ueda et al. 2005) for the Si edge. We modeled
each edge by fitting a function of the form
\begin{eqnarray}
F_{\lambda} = \left \{
\begin{array}{lr}
F^{\rm c}_{\lambda} & {\rm for} \lambda > \lambda_{\rm edge}  \\
F^{\rm c}_{\lambda} \exp\left[-A \left( \frac{\lambda}{\lambda_{\rm edge}} \right)^3 \right]   
& {\rm for} \lambda \leq \lambda_{\rm edge} 
\end{array} \right.
\end{eqnarray}

We adjusted the specific wavelength range that we used in the fits for
each source because of small variations in the intrinsic continuum
spectra of the sources. However, between the multiple observations of
a given a source, we froze the range that we analyzed.  Specifically,
for GX~17$+$2, we used $5.8-7.2~{\rm \AA}$ for Si and $8.6-10.5~{\rm
\AA}$ for Mg. For 4U~1705$-$44, we chose a range of $5.7-7.5~{\rm
\AA}$ for Si and 8.5$-$10.3~${\rm \AA}$ for Mg. Finally, for
4U~1728$-$34, we used $5.9-7.3~{\rm \AA}$ for Si and $8.5-10.3~{\rm
\AA}$ for Mg.  Sample spectra for all three sources in these
wavelength ranges are shown in Figures 3 and 4.

We fit the MEG and HEG data both separately and simultaneously to
detect any possible systematic differences. For the data taken in the
TE mode, the slopes obtained by fitting power-law functions to the
continuum of both gratings were in good agreement with each other.
Note that the pile-up fraction in one of the observations of
4U~1705$-$44 (Obsid: 1923) was nearly 15\% around the Si edge in the
MEG spectrum and this result was excluded from the final analysis. In
the CC mode observations, however, there were more than 3$\sigma$
differences in the power-law indices obtained for the HEG and MEG
data. This is thought to be due to calibration uncertainties in this
mode. Because of this, only MEG data were used for all of the fits (of
the CC mode data) due to the larger photon collecting area of this
grating compared to the HEG in the wavelength region of interest
($5.5-15.5~{\rm \AA}$).

In addition to a power-law continuum and the function modeling the
edges, we allowed for a number of emission and absorption line
features that appear in each spectral region and whose positions are
fixed. For consistency, we included the same number of absorption and
emission lines in the spectral fits of each observation of each
source, regardless of the apparent presence or absence of a feature in
that spectrum, as was done by Juett et al.\ (2006). When the features
were not detectable or were not well-constrained in a given
observation, we report a zero flux for the lines. A negative amplitude
corresponds to an absorption feature.

Around the Si absorption edge, we fixed the wavelengths of the five
most prominent features in the wavelength range for each source using
Chandra's atomic lines database ATOMDB\footnotemark[3]. The most
prominent features in this region are the two Si XIV and three Si XIII
atomic lines. In Figure~3, we show three sample spectra from each
source around the Si edge, along with the fitted power-law continuum,
the edge, and the line features. Table~2 shows the calculated fluxes
of the emission lines for each observation.

Similarly, in the wavelength range around the Mg edge, we included
Mg~XI, Fe~XXI, and Ne~X emission features, which are the
strongest. The Mg~XI features at 9.16~{\rm\AA}, 9.22~{\rm \AA}, and
9.31~{\rm \AA} are part of a He-like triplet formed by the resonance,
intercombination, and forbidden transitions (Porquet et al.\ 2001).
We show in Figure~4 the results of our fits to the grating spectra
obtained from one of the observations of each source. We report the
corresponding line fluxes in Table~3 and report our findings for the
Si and Mg edges in the next section.

\footnotetext[1]{http://asc.harvard.edu/ciao/threads/}
\footnotetext[2]{http://space.mit.edu/cxc/analysis/findzo/Data/findzo.sl}
\footnotetext[3]{http://cxc.harvard.edu/atomdb/WebGUIDE/}

\section{Results and Discussion}

We used the Chandra grating observations of GX~17$+$2, 4U~1705$-$44,
and 4U~1728$-$34 to measure absorption coefficients for the Mg and Si
edges. In Tables 2 and 3, we also report the fluxes of spectral lines
that were measured in each observation around the Si and Mg edges,
respectively. We then used the absorption coefficients along with the
cross-sections from Gould \& Jung (1991) to find the column densities
for each element. We converted the Mg and Si column densities to
hydrogen column densities, $N_{\rm H}$, using the ISM abundances given
by Wilms, Allen, \& McCray (2000). We report all of the column
densities in Table~4. We then averaged the results obtained with each
absorption edge over the observations, which makes it possible to
discriminate any possible systematic differences between the column
densities obtained from the different edges.

For GX~17$+$2, the $N_{\rm H}$ values derived from the Mg edge ranged
from $(2.01-2.77) \times 10^{22}~{\rm cm}^{-2}$, while the values
derived from the Si edge ranged from $(1.03-2.69) \times 10^{22}~{\rm
cm}^{-2}$ for the three observations. Averaging over the column
densities derived from the Mg edge yields $(2.38 \pm 0.12) \times
10^{22}~{\rm cm}^{-2}$, while over those derived from the Si edge results
in $(1.78 \pm 0.09) \times 10^{22}~{\rm cm}^{-2}$. We note that the
lower $N_{\rm H}$ value derived from the Si edge can be attributed to
the small edge measured in one single observation and that the
remaining five measurements are consistent with each other to within
2$\sigma$. 

4U~1705$-$44 showed a somewhat smaller scatter with equivalent $N_{\rm
H}$ ranging from $(2.08-2.59) \times 10^{22} {\rm cm}^{-2}$ from the
Mg edge measurements and $(2.95 - 4.12) \times 10^{22} {\rm cm}^{-2}$
from the Si edge measurements. Averaging over the column densities
derived from the Mg edge yields $(2.44 \pm 0.09) \times 10^{22}~{\rm
cm}^{-2}$, and $(3.64 \pm 0.12) \times 10^{22}~{\rm cm}^{-2}$ from the
Si edge. Because the difference between these two measurements are
larger than that expected from statistical uncertainties, we treat it
as a systematic difference, potentially caused by calibration errors
or by anomalous abundances.

Finally, 4U~1728$-$34 had equivalent hydrogen column densities in the
range $(2.39-8.23) \times 10^{22}~{\rm cm}^{-2}$ in individual edge
measurements. However, the highest of these values, corresponding to
the Silicon edge in observation 2748 (N[Si]$_{\rm H} = (8.23 \pm 0.29)
\times 10^{22}~{\rm cm}^{-2}$), arises from a previously reported Si
overabundance that is thought to be caused from calibration
problems in that specific observation (D'A{\'{\i}} et al.\ 2006). As a
result, we excluded this measurement when calculating the average
hydrogen column density for this source. We again calculated separate
average values for each edge: we obtained $N_{\rm H} = (2.49 \pm 0.14)
\times 10^{22}~{\rm cm}^{-2}$ from the Mg edge and $(4.27 \pm 0.14)
\times 10^{22}~{\rm cm}^{-2}$ from the Si edge.

We detect a systematic difference between the $N_{\rm H}$ values
obtained from Mg and Si edges in all of the observations. These could
naturally be attributed to non-standard ISM abundances or to anomalous
intrinsic abundances of the sources. However, we also observed that
the scatter between the absorption coefficients at the Si edge is also
much larger between the different observations of each source. Noting
that the response of the detector suffers from a very large
discontinuity at the Si edge but is smooth across the Mg edge, we
investigated whether these differences could be due to small
inaccuracies in the detector response across the Si edge.  In
particular, we explored whether there is any dependence of this
systematic variation on the source count rate during the observations.
We found that when the count rates of the sources are on average
higher ($\sim 30$~cts~s$^{-1}$), the N$_{\rm H}$ values found using
the two edges become consistent with each other to within $2\sigma$,
and that the discrepancy between the two values increases as the
source count rate decreases. This is especially evident in the two
spectra of 4U~1705$-$44, where one of the two spectra of this source
was obtained when the source count rate was 38.6~cts~s$^{-1}$ while
the second observation was performed during a smaller count rate epoch
(12.6~cts~s$^{-1}$). We note that between these observations the
N$_{\rm H}$ value found from the Mg edge does not change
significantly, while the value found from the Si edge increases and
the difference between the two measurement becomes as much as
10$\sigma$.

Given that the intrinsic metal abundance of the X-ray binary or the
metal abundances in the ISM is unlikely to change significantly for
the same source within 4 years, we conclude that the difference might
arise because of the presence of the significant intrumental Si edge,
where small uncalibrated residuals of this structure affect our edge
measurements. Because of this, we will use $N_{\rm H}$ measurements
from the Mg edges as the preferred values.

We have also compared our measurements of hydrogen column densities
towards these three sources determined from the overall suppression of
low-resolution continuum spectra.

Previous estimates of the column density towards GX~17$+$2 were made
from modeling the continuum spectra obtained with ROSAT observations.
Predehl \& Schmitt (1995) found $N_{\rm H}$ values to the source in
the range $(1.79-2.34) \times 10^{22}$~cm$^{-2}$, assuming a blackbody,
thermal bremsstrahlung or a power-law model for the continuum.  These
values span the range obtained from the Mg and Si edge measurements
and are thus in agreement with our results.

Two estimates of the column density towards 4U~1705$-$44 exist based
on observations with different satellites and using different
continuum models. Predehl \& Schmitt (1995) reported a range of
hydrogen column densities towards towards the source using ROSAT
observations. Continuum models included a blackbody, thermal
bremsstrahlung or a power-law, as in the case of GX~17$+$2, and
resulted in a range $N_{\rm H} = (1.23 - 1.63) \times 10^{22}~{\rm
cm}^{-2}$. Barret \& Olive (2002), on the other hand, reported a
column density $N_{\rm H} = 2.4 \times 10^{22}~{\rm cm}^{-2}$ by
fitting a Comptonized blackbody model to the RXTE Proportional Counter
Array data. The former values are lower than the model-independent
column densities we find here, while the latter value is consistent
within $1\sigma$ with our results when only the Mg column is taken
into account.

Observations of 4U~1728$-$34 by ROSAT, BeppoSax, and Chandra
observatories have also yielded a range of estimates for the hydrogen
column density towards this source when the data were fit with
different continuum models. Schulz (1999) analyzed the ROSAT data
using power-law and blackbody continuum models, which yielded hydrogen
column densities $N_{\rm H} = (3.44 \pm 0.3) \times 10^{22}~{\rm
cm}^{-2}$ and $N_{\rm H} = (2.89 \pm 0.22) \times 10^{22}~{\rm
cm}^{-2}$, respectively. BeppoSAX data were fit with a Comptonized
blackbody model to give $N_{\rm H} = (2.73 \pm 0.05) \times
10^{22}~{\rm cm}^{-2}$ (Piraino, Santangelo, \& Kaaret 2000).
Finally, the spectrum obtained with Chandra HETG was modeled with a
Comptonized blackbody to give 2.61$^{+.06}_{-.07} \times
10^{22}~\rm{cm}^{-2}$ (D'A{\'{\i}} et al.\ 2006). Note that while
4U~1728$-$34 has also been observed with the RXTE Proportional Counter
Array and Integral, the hydrogen column density was not well
constrained from these observations. All but the ROSAT measurements
are consistent with the column density of $N_{\rm H}=(2.49 \pm 0.14)
\times 10^{22}~{\rm cm}^{-2}$ measured from the Mg edge.

\acknowledgments
We thank Adrienne Juett for numerous helpful discussions and
suggestions on the analysis of the grating spectra as well as her
comments on the manuscript. We also thank Michael Nowak for his help
with Chandra HETG. This work was supported by NSF grant AST~07-08640.

\clearpage

\begin{figure*} \centering
\includegraphics[angle=0, scale=0.70]{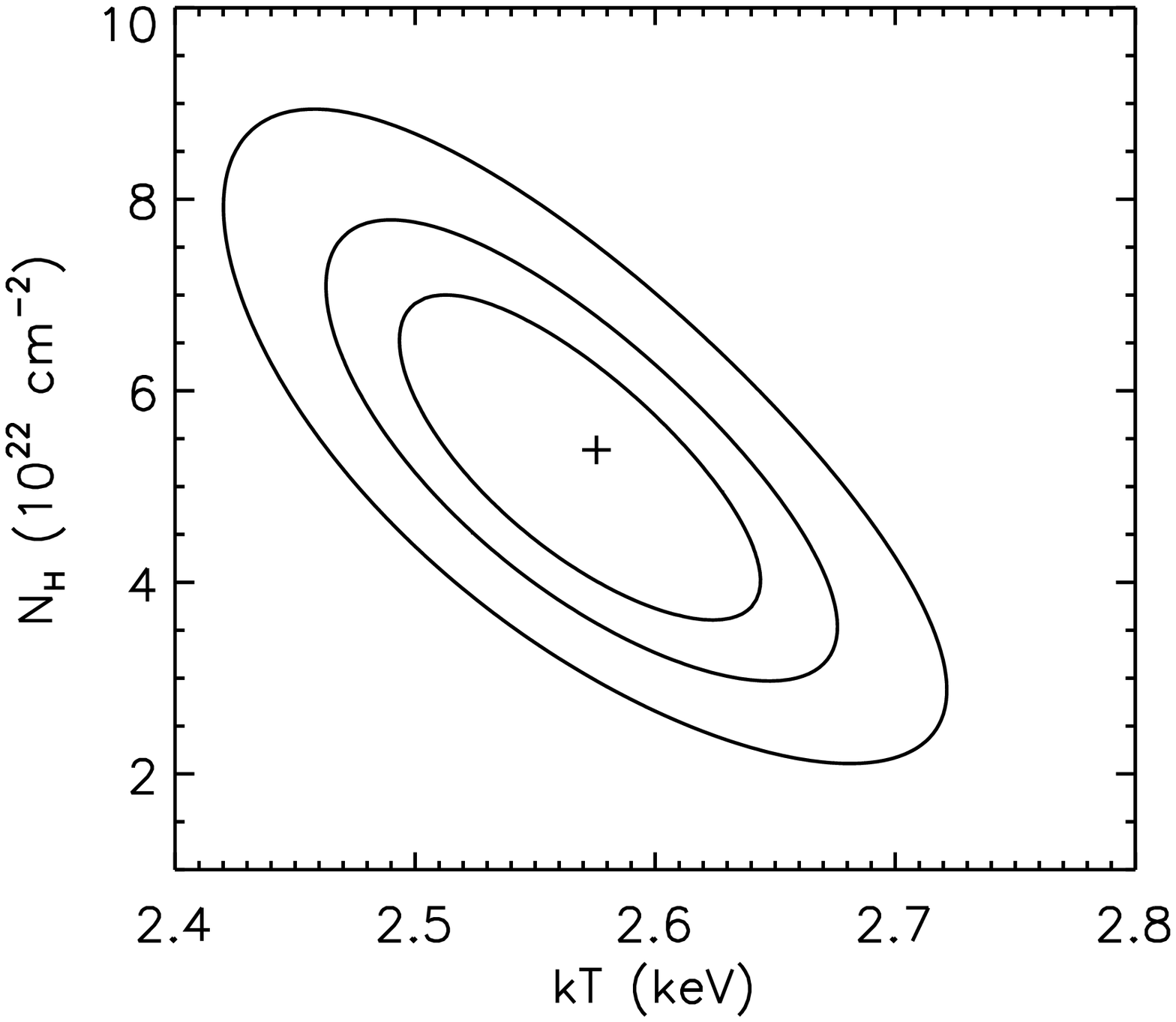}
\caption{ The correlation between column density and temperature is
shown for 4U~1728$-$34 as found for one of the photospheric radius
expansion bursts from an RXTE observation. The three contours show the
1$\sigma$, 2$\sigma$, and 3$\sigma$ confidence levels. }
\label{correlation} 
\end{figure*}

\begin{figure*}
\centering
   \includegraphics[scale=0.70]{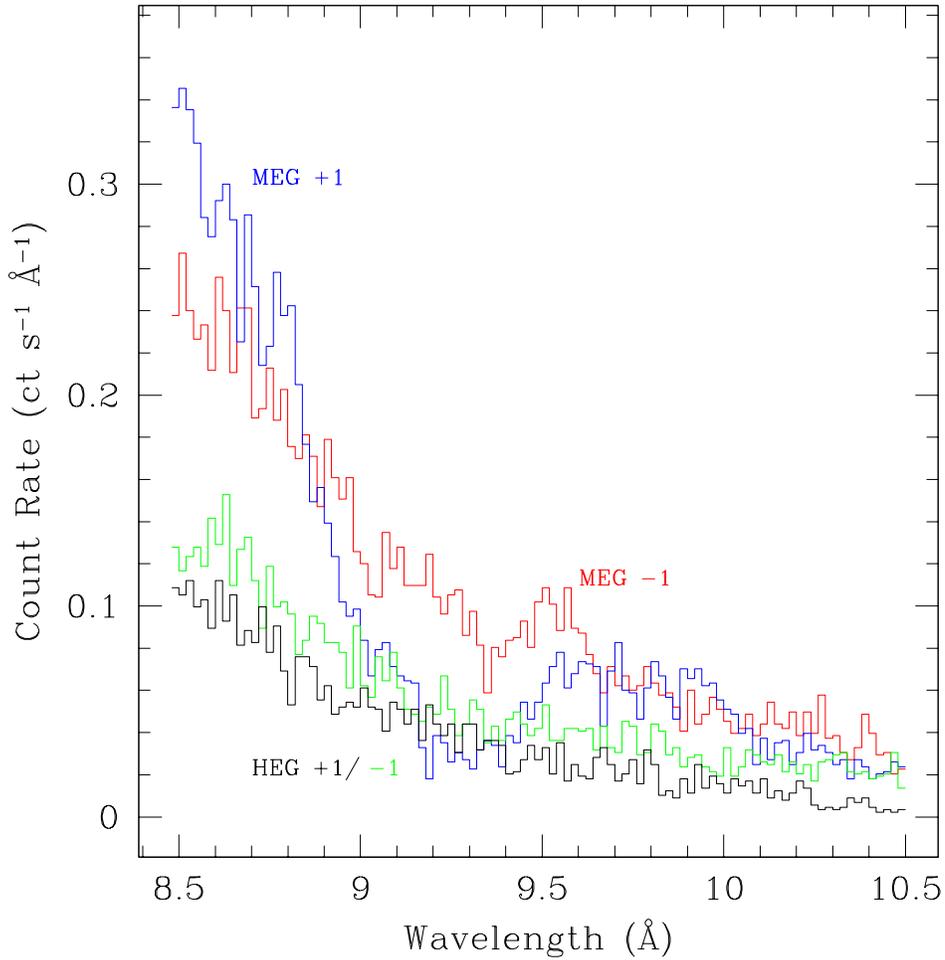} \caption{A sample spectrum of
   4U~1728$-$34 that shows the $+1$ and $-1$ orders of both MEG and
   HEG gratings. In addition to intrinsic differences between the
   effective areas of the HEG and the MEG gratings, there is also a
   chip gap at the MEG $+1$ order aroung 9.5~$\AA$ which causes a
   significant reduction in the counts in this region.}
\label{effarea}
\end{figure*}

\begin{figure*}
\centering
   \includegraphics[scale=0.70]{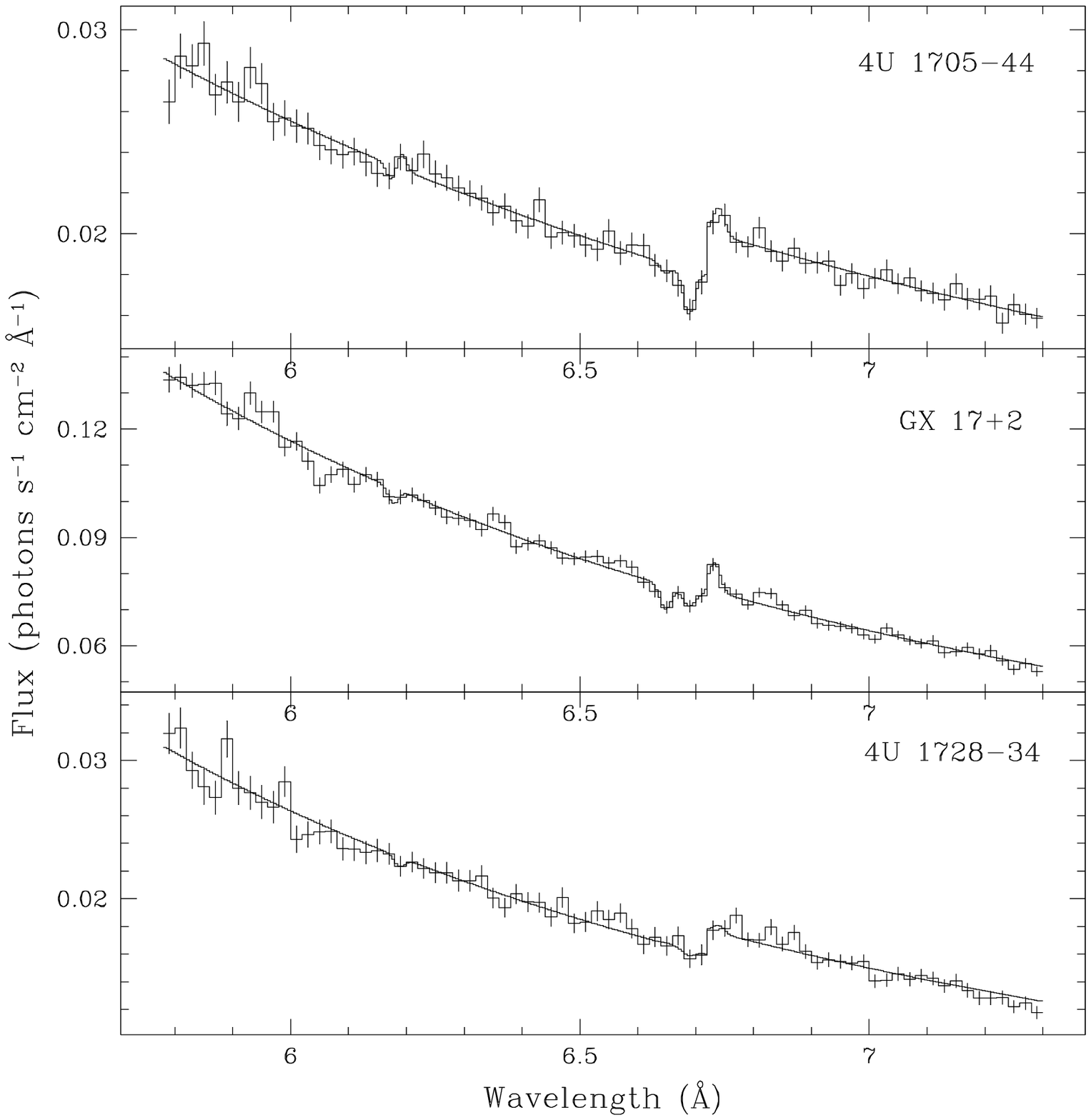}
   \caption{The fitted spectra for the Silicon edge. For
     4U~1705$-$44, GX~17$+$2, and 4U~1728$-$34, obsids 5500, 4564, and
     6568, are shown, respectively. Details of these fits can be seen in 
     Table~2. }
\label{Siedge}
\end{figure*}

\begin{figure*}
\centering
\includegraphics[scale=0.70 ]{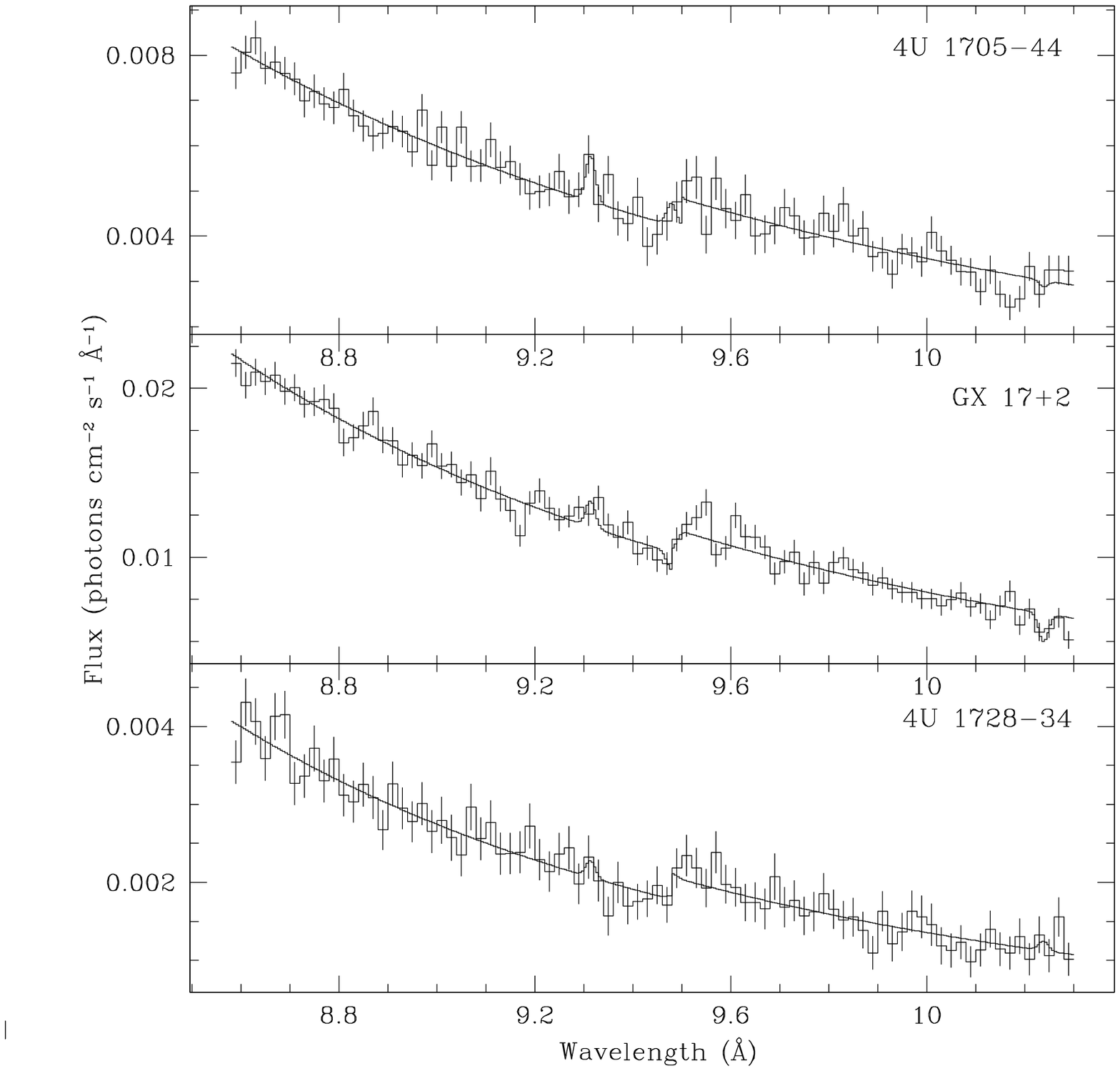}
\caption{The fitted spectra for the Magnesium
edge. For 4U~1705$-$44, GX~17$+$2, and 4U~1728$-$34, obsids 5500, 4564, 
and 6568, are shown, respectively. Details of these fits can be seen in 
Table~3. }
\label{Mgedge}
\end{figure*}

\begin{deluxetable}{lccccc}
\tablecolumns{4}
\tablewidth{460.0pt}
\tablecaption{List of Chandra Observations by Source Used in this Study}
\tablehead{Source & Obsid & Mode & Date & Exp. Time (ks) & Average Count Rate (ct~s$^{-1}$)}
\startdata
GX~17$+$2  & 4564  & CC &   2004 July 11 & 30.18 & 30.43   \\
           & 6629  & CC &   2006 May 10  & 23.68 & 31.52  \\
           & 6630  & CC &   2006 Aug 19  & 24.08 & 29.82  \\
4U~1705$-$44 & 5500  & TE & 2001 July 01 & 25.13 & 12.62 \\
             & 1923  & TE & 2005 Oct 26  & 27.25 & 38.60 \\
4U~1728$-$34 & 2748  & TE & 2002 Mar 04  & 10.00 & 7.29  \\
             & 6568  & CC & 2006 July 17 & 49.49 & 6.53  \\
             & 6567  & CC & 2006 July 18 & 151.84 & 7.84 \\
             & 7371  & CC & 2006 July 22 & 39.71 & 4.53 \\
\enddata
\end{deluxetable}

\begin{deluxetable}{lccccccc}
\tablecolumns{8}
\tablewidth{460.0pt}
\tablecaption{Fluxes of Spectral Lines around the Si Edge$^1$} 
\tablehead{
 &      & \multicolumn{5}{c}{Flux ($10^{-5}$ photons $\rm cm^{-1}$ $\rm sec^{-1}$ $\rm\AA^{-1}$)} \\
 &      & Si XIV$_a^{2}$          & Si XIV$_b$    & Si XIII$_c$   &  Si XIII$_d$  & Si XIII$_e$ \\
 Source   & Obsid & 6.1804~$\rm\AA$ & 6.1858~$\rm\AA$ & 6.6479~$\rm\AA$ & 6.6882~$\rm\AA$ & 6.7403~$\rm\AA$}
\startdata
GX~17$+$2 & 4564 &-16.0$\pm$5.99$^{1}$& $--$           & -17.3$\pm$4.8   & -10.1$\pm$4.8 & 15.8$\pm$4.6 \\
          &6629 & $--$                & $--$           & $--$            & -13.1$\pm$9.0 & 26.1$\pm$8.5\\
          &6630 & -11.7$\pm$6.85      & $--$           & -11.7$\pm$5.8   & $--$           &  9.64$\pm$5.38\\
4U~1705$-$44 &1923 & $--$ & $--$ & $--$ & -11.03$\pm$5.46 & $--$     \\
          &5500 &  -3.43$\pm$1.76     & 3.93$\pm$1.69  &$--$             & -5.59$\pm$1.46 & 5.42$\pm$1.44 \\
4U~1728$-$34    &2748 & $--$          & 1.66$\pm$1.43  & 2.07$\pm$1.20   & 3.34$\pm$1.16  & $--$ \\
          &6567 & 2.06$\pm$1.99       & 2.80$\pm$1.89  &$--$             & -2.13$\pm$1.66 & 7.36$\pm$1.60\\
          &6568 &  $--$               &$--$            & $--$            & $--$           & $--$ \\
          &7371 &  -5.07$\pm$2.12     & 4.69$\pm$2.04  & -4.81$\pm$1.8   & $--$           & 3.87$\pm$1.76\\
\enddata
\tablenotetext{1}{Errors correspond to 1$\sigma$ statistical uncertainties.}
\tablenotetext{2}{Transitions from $^{a} 4\rightarrow1$ $^{b} 3\rightarrow1$ $^{c} 7\rightarrow1$ $^{d} 5\rightarrow1$ $^{e} 2\rightarrow1$}
\end{deluxetable}

\begin{deluxetable}{lccccccc}
\tablecolumns{8}
\tablewidth{500.0pt}
\tablecaption{Fluxes of Spectral Lines around the Mg Edge$^1$} 
\tablehead{
         & & \multicolumn{4}{c}{Flux ($10^{-5}$ photons $\rm cm^{-1}$
         $\rm sec^{-1}$ $\rm\AA^{-1}$)} \\ & & Mg XI$_a^{2}$ & Mg
         XI$_b$ & Mg XI$_c$ & Fe XXI & Ne X \\ Source & Obsid & 9.1687~$\rm\AA$ 
         & 9.2267~$\rm\AA$  & 9.3143~$\rm\AA$  & 9.4797~$\rm\AA$  & 10.2385~$\rm\AA$  \\}
\startdata
GX~17$+$2 &4564 &$--$          & $--$          &3.95$\pm$2.17    & 2.95$\pm$2.17     & -4.37$\pm$1.58\\
          &6629 &$--$          & $--$          &$--$             & $--$              & $--$ \\
          &6630 &$--$          & $--$          &  $--$           & $--$              & -4.98$\pm$1.91\\
4U~1705$-$44&1923 &$--$        & $--$          &  $--$           & $--$             &$--$ \\
	  &5500 &$--$          & 2.72$\pm$1.11& 1.52$\pm$1.22  & $--$              & $--$ \\
4U~1728$-$34&2748 &$--$        & $--$        & $--$            & $--$              & $--$ \\
          &6567 &1.51$\pm$0.71 & 1.69$\pm$0.71  & $--$            & -0.93$\pm$0.57      &  $--$\\
          &6568 &$--$          & $--$          &$--$            & $--$              & $--$  \\
          &7371 &2.90$\pm$0.94 &1.54$\pm$0.84 & $--$             & $--$           & -0.61$\pm$0.56  \\
\enddata
\tablenotetext{1}{Errors correspond to 1$\sigma$ statistical uncertainties.}
\tablenotetext{2}{Transitions from $^{a} 7\rightarrow1$ $^{b} 6\rightarrow1$ $^{c} 2\rightarrow1$}

\end{deluxetable}
\begin{deluxetable}{lcccccc}
\tablecolumns{7}
\tablewidth{460.0pt}
\tablecaption{Column Densities from Mg and Si Edges$^1$}
\tablehead{
Source     & Obsid & N$_{\rm Mg}$& N[Mg]$_{\rm H}$ & N$_{\rm Si}$ & N[Si]$_{\rm H}$ & $\langle N_{\rm H} \rangle$ \\
           &       & $10^{17}~{\rm cm}^{-2}$ & $10^{22}~{\rm cm}^{-2}$ & $10^{17}~{\rm cm}^{-2}$ 
           & $10^{22}~{\rm cm}^{-2}$ & $10^{22}~{\rm cm}^{-2}$ }
\startdata
GX~17$+$2    & 4564  & 5.06$\pm$0.41  & 2.01$\pm$0.16 & 1.92$\pm$0.26 & 1.03$\pm$0.14  &  \\
             & 6629  & 6.95$\pm$1.08  & 2.77$\pm$0.43 & 5.01$\pm$0.44 & 2.69$\pm$0.23  &  \\
             & 6630  & 6.96$\pm$0.44  & 2.77$\pm$0.18 & 4.33$\pm$0.29 & 2.32$\pm$0.16  &  2.38$\pm$ 0.12$^{2}$\\
4U~1705$-$44 & 1923  & 6.50$\pm$0.26  & 2.59$\pm$0.10 & 5.49$\pm$0.36 & 2.95$\pm$0.19  &  \\
             & 5500  & 5.23$\pm$0.41  & 2.08$\pm$0.16 & 7.66$\pm$0.30 & 4.12$\pm$0.16  &  2.44$\pm$ 0.09 \\
4U~1728$-$34 & 2748  & 6.38$\pm$0.81  & 2.54$\pm$0.32 & 15.3$\pm$0.55 & 8.23$\pm$0.29  &  \\
             & 6567  & 6.08$\pm$0.51  & 2.42$\pm$0.20 & 8.36$\pm$0.33 & 4.49$\pm$0.18  &  \\
             & 6568  & 6.80$\pm$0.85  & 2.71$\pm$0.34 & 7.32$\pm$0.47 & 3.93$\pm$0.25  &  \\
             & 7371  & 6.01$\pm$0.99  & 2.39$\pm$0.39 & 7.22$\pm$0.98 & 3.88$\pm$0.53  &  2.49$\pm$0.14\\

\enddata
\tablenotetext{1}{Errors correspond to 1-$\sigma$ statistical uncertainties.}
\tablenotetext{2}{The average value for all observations per source for the Mg edge.} 

\end{deluxetable}


\begin{thebibliography}

\bibitem[Barret \& Olive(2002)]{2002ApJ...576..391B} Barret, D., \& Olive, 
J.-F.\ 2002, \apj, 576, 391 

\bibitem[D'A{\'{\i}} et al.(2006)]{2006A&A...448..817D} D'A{\'{\i}}, A., 
et al.\ 2006, \aap, 448, 817 

\bibitem[Damen et al.(1990)]{1990A&A...237..103D} Damen, E., Magnier,
E., Lewin, W.~H.~G., Tan, J., Penninx, W., \& van Paradijs, J.\ 1990,
\aap, 237, 103

\bibitem[Durant \& van Kerkwijk(2006)]{2006ApJ...650.1082D} Durant, M., 
\& van Kerkwijk, M.~H.\ 2006, \apj, 650, 1082 

\bibitem[Galloway et al.(2003)]{2003ApJ...590..999G} Galloway, D.~K., 
Psaltis, D., Chakrabarty, D., \& Muno, M.~P.\ 2003, \apj, 590, 999

\bibitem[Galloway et al.(2008)]{2006astro.ph..8259G} Galloway, D.~K.,
Muno, M.~P., Hartman, J.~M., Psaltis, D., \& Chakrabarty, D.\ 2008,
ApJS, in press (arXiv:astro-ph/0608259)

\bibitem[Gould \& Jung(1991)]{1991ApJ...373..271G} Gould, R.~J., \& Jung, 
Y.-D.\ 1991, \apj, 373, 271 

\bibitem[Juett et al.(2004)]{2004ApJ...612..308J} Juett, A.~M., Schulz, 
N.~S., \& Chakrabarty, D.\ 2004, \apj, 612, 308 

\bibitem[Juett et al.(2006)]{2006ApJ...648.1066J} Juett, A.~M., Schulz, 
N.~S., Chakrabarty, D., \& Gorczyca, T.~W.\ 2006, \apj, 648, 1066 

\bibitem[Lattimer \& Prakash(2001)]{2001ApJ...550..426L} Lattimer,
J.~M., \& Prakash, M.\ 2001, \apj, 550, 426

\bibitem[Lattimer \& Prakash(2007)]{2007PhR...442..109L} Lattimer,
J.~M., \& Prakash, M.\ 2007, \physrep, 442, 109

\bibitem[Lee et al.(2001)]{2001AJ....122.3136L} Lee, J.-W., Carney, B.~W., 
Fullton, L.~K., \& Stetson, P.~B.\ 2001, \aj, 122, 3136 

\bibitem[Lewin et al.(1993)]{1993SSRv...62..223L} Lewin, W.~H.~G., van 
Paradijs, J., \& Taam, R.~E.\ 1993, Space Science Reviews, 62, 223 

\bibitem[L{\'o}pez-Corredoira et al.(2002)]{2002A&A...394..883L} L{\'o}pez-Corredoira, 
M., Cabrera-Lavers, A., Garz{\'o}n, F., \& Hammersley, P.~L.\ 2002,
\aap, 394, 883

\bibitem[{\"O}zel(2006)]{2006Natur.441.1115O} {\"O}zel, F.\ 2006, \nat, 
441, 1115

\bibitem[Ozel et al 2008]{} \"Ozel, F., G\"uver, T., \& Psaltis, D., ApJ, 
submitted

\bibitem[Paczynski \& Stanek(1998)]{1998ApJ...494L.219P} Paczynski, B., \& 
Stanek, K.~Z.\ 1998, \apjl, 494, L219 

\bibitem[Piraino et 
al.(2000)]{2000A&A...360L..35P} Piraino, S., Santangelo, A., \& Kaaret, P.\ 2000, \aap, 360, L35 

\bibitem[Predehl \& Schmitt(1995)]{1995A&A...293..889P} Predehl, P., \& Schmitt, 
J.~H.~M.~M.\ 1995, \aap, 293, 889 

\bibitem[Porquet et al.(2001)]{2001A&A...376.1113P} Porquet, D., Mewe, R., 
Dubau, J., Raassen, A.~J.~J., \& Kaastra, J.~S.\ 2001, \aap, 376, 1113 

\bibitem[Schulz(1999)]{1999ApJ...511..304S} Schulz, N.~S.\ 1999, \apj, 511, 
304 

\bibitem[Ueda et al.(2005)]{2005ApJ...620..274U} Ueda, Y., Mitsuda, K., 
Murakami, H., \& Matsushita, K.\ 2005, \apj, 620, 274

\bibitem[van Paradijs(1978)]{1978Natur.274..650V} van Paradijs, J.\ 1978, 
\nat, 274, 650 

\bibitem[Wilms et al.(2000)]{2000ApJ...542..914W} Wilms, J., Allen, A., 
\& McCray, R.\ 2000, \apj, 542, 914 


\end{thebibliography}
\end{document}